\documentclass{article}
\usepackage[utf8]{inputenc}
\usepackage{authblk}
\usepackage[dvips]{graphicx}
\graphicspath{{noiseimages/}}
\usepackage{xcolor}
\usepackage[charter]{mathdesign}
\let\mathcal\undefined
\DeclareMathAlphabet{\mathcal}{OMS}{cmsy}{m}{n}
\usepackage[T1]{fontenc}
\usepackage{tikz}
\usepackage{empheq}
\newcommand*\widefbox[1]{\hspace{2em}\fbox{\hspace{0.5em}#1\hspace{0.5em}}}
\usepackage{MnSymbol}
\usepackage{pgfplots}
\pgfplotsset{compat=1.7}
\usetikzlibrary{intersections, pgfplots.fillbetween}
\usetikzlibrary{decorations.pathmorphing}
\usetikzlibrary{arrows.meta}
\usepackage[mode=buildnew]{standalone}
\usepackage[toc,page]{appendix}
\usepackage{amsmath}
\usepackage{float}
\usepackage{comment}
\usepackage{bm}
\usepackage{a4wide}
\usepackage{bm}
\newcommand{\XI}{\boldsymbol{\mathbf{\zeta}}}

\usepackage[english]{babel}
\usepackage{hyperref}
\usepackage{caption}
\AtBeginDocument{}
\definecolor{refcolor}{HTML}{CD2600}
\definecolor{tablecolor}{HTML}{373641} 
\definecolor{urlcolor}{HTML}{E12900} 
\hypersetup{pdfstartview=FitH,  linkcolor=refcolor,urlcolor=urlcolor, colorlinks=true,citecolor=refcolor}
\usepackage[page]{appendix}

\newcommand{\w}{\omega}

\author[1,2]{D. V. Diakonov\footnote{\tt dmitrii.dyakonov@phystech.edu}}
\author[1,2]{K. V. Bazarov\footnote{\tt bazarov.kv@phystech.edu}}
\affil[1]{Moscow Institute of Physics and Technology, Institutskii per. 9, 141700, Dolgoprudny, Russia}
\affil[2]{NRC ”Kurchatov Institute”, 123182, Moscow, Russia}
\title{\textcolor{black}{Thermal loops in the accelerating frame}}
\begin{document}

\numberwithin{equation}{section}

\maketitle
We consider the conformal scalar field theory with $\lambda \phi^4$ self-interaction in Rindler and Minkowskian coordinates at finite temperature planckian distribution for the exact modes. The solution of the one-loop Dyson-Schwinger equation is found to the order in $\lambda^{3/2}$. Appearance of the thermal (Debye) mass is shown. Unlike the physical mass, the thermal one gives a gap in the energy spectrum in the quantization in the Rindler coordinates. The difference between such calculations in Minkowski and Rindler coordinates for the exact modes is discussed. It is also shown that states with a temperature lower than the Unruh one are unstable. It is proved that for the canonical Unruh temperature the thermal mass is equal to zero. The contribution to the quantum average of the stress-energy tensor is also calculated, it remains traceless even in the presence of the thermal mass.
\newpage 
{
  \hypersetup{linkcolor=tablecolor}
  \tableofcontents
}
\newpage
\section{Introduction}
Quantum field theory is a well developed instrument to consider wide range of effects in various fields of fundamental physics. An interesting subtopic with intriguing results is quantum field theory in curved spaces-times. For example, not long after pioneering works \cite{Fulling:1972md,Davies:1974th,Hawking:1975vcx,Unruh:1976db,Gibbons:1977mu, Unruh:1983ms}, it became clear that quantum field theory in the static gravitational backgrounds bordered by Killing horizons possesses certain \emph{thermal} properties. A great stimulus for a systematic investigation of the quantum field theory in curved space-times is a relationship between gravity and thermodynamics.

These \emph{thermal} effects lead to the famous effects such as the black hole radiation \cite{Hawking:1975vcx} or Bekenstein-Hawking entropy \cite{Bekenstein:1972tm,Bekenstein:1973ur,Hawking:1975vcx,Gibbons:1977mu}. The latter is entropy proportional to the area of the horizon and is conceptually different from the usual entropy of the thermal gas in Minkowski space-time, which is proportional to the space-time volume. For more details see, e.g. \cite{Solodukhin:2011gn}.

It seems fair to say that the canonical (Unruh, Hawking and Gibbons-Hawking) temperatures play central role in topics under discussion. Corresponding states are defined by a geometry of the given space-time and are in thermal equilibrium with the geometry. But what about the thermal states with arbitrary temperatures in static spaces with horizons? It is a good moment to noting
that the thermalization process in curved space-times in general is far from being well understood, e.g. \cite{Akhmedov:2021rhq}. So, thermal states with an arbitrary temperature may shed some light on the general thermal properties in curved spaces. From a technical point of view, thermodynamic quantities are usually expressed in terms of derivatives of the partition function with respect to the temperature, therefore it is necessary to know the value of the free energy not only for the canonical temperature but for an arbitrary temperature (at least for the temperatures which are close to the canonical one). Finally, it is shown that thermal states under consideration do not possess the same essential thermal properties as those in Minkowksi space-time: see e.g. \cite{Akhmedov:2020qxd, Akhmedov:2020ryq, Anempodistov:2020oki, Akhmedov:2021cwh, Bazarov:2021rrb}. Therefore, we find it interesting to check the other thermal properties of the QFT in space-times with horizons. Namely, to start with we consider the Rindler wedge of the Minkowksi space-time, as the model example.

We propose to consider a thermal phenomenon so called the generation of the Debye screening in plasma. Usually, it is discussed in the context of QED, where the photon acquires an effective mass when it propagates through a neutral and hot plasma of electrons and positrons \cite{Ahmed:1991mz,Arnold:1995bh, Masood:2018uuw}. The main question addressed in this article is: Will the thermal equilibrium states in the Rindler space-time generate the Debye mass? And what will be the physical manifestations of such a mass? We also compare these properties with the corresponding properties in the Minkowski space-time.

Loop calculations in QED in curved (curvilinear) coordinates are cumbersome. Instead, we choose to study a toy model to demonstrate the basic properties. This toy model is conformal scalar field theory with self-interaction:
\begin{align}
\label{action}
Z=\int D\phi \  exp\Bigg[- \int_0^\beta d\tau \int d^3 x \Big(-\frac{1}{2}\partial_{\mu} \phi(X)\partial^{\mu} \phi(X)+\frac{\lambda}{4!}\phi^4(X)+\XI R \phi^2(X)\Big)\Bigg].
\end{align}
Here $X \equiv (\tau, \vec{x})$ and $\beta=\frac{1}{T}$ is an inverse temperature. Note that we work in the following signature $(-,-,-,-)$, so the first term in \eqref{action} has opposite sign.\footnote{We choose such a seemilyly inconvenient signature because after analytical continuation $\tau=-it$ the metric has convenient $(+,-,-,-)$ signs.} Conformal theory corresponds to the case, when $\XI=\frac16$. Note that Maxwell theory in four dimensions is conformally invariant.

The paper is organized as follows: in the Sec. \ref{debmassMINsec} we review well-known results about the thermal quantum field theory in the Minkowski space-time. In the Sec. \ref{M1}-\ref{M3} we discuss straightforward summation of tadpole diagrams, Dyson equation and the origin of non-pretrubative contributions to the thermal mass. In the Sec. \ref{setMIN} we show that the contributions to the stress–energy tensor from the thermal and the physical masses are different. In Sec. \ref{gensecRindler} we review definitions and basic properties of the bare thermal propagator in the Rindler coordinate. In the Sec. \ref{massRindler} we discuss the regularizated value of the propagator at the coincident points and straightforward summation of the tadpole diagrams. In this section, we show that the thermal mass gives a gap in the energy spectrum, although the physical mass does not. In the Sec. \ref{DS Rind} and \ref{32contributionMIN} we discuss Dyson equation and show how the non-analytical $\lambda^\frac{3}{2}$ term in the thermal mass can be obtained directly. In the Sec. \ref{SecsetRin} we calculate the stress–energy tensor and compare it with the Minkowskian one. In the Sec. \ref{Highdim} we discuss loop corrections in the higher dimensions. We put all the relevant details of the calculations into the Appendix.
\section{Thermal mass in Minkowski space-time }
\label{debmassMINsec}
In this section, we briefly recall ideas and details of the calculation in quantum field theory in Minkowski space-time with the action \eqref{action} at finite temperature. We also discuss thermal (Debye) mass and its contribution to the stress-energy tensor. The major part of this section consists of the well-known results (good reviews can be found, for example, in \cite{kapusta_gale_2006,Laine:2016hma}). We recall these classical results to underline the differences and similarities of the thermal effects in the Minkowski and Rindler coordinates for the exact modes and to clarify the logic of the regularization scheme. We start with the standart Feynman thermal (or Matsubara) propagator in Minkowski space-time:
\begin{align}
\label{green0MIN}
\langle T \phi(X_1) \phi(X_2) \rangle \Big|_{\lambda=0}=G_{\lambda^{0}}(X_1,X_2) \equiv \frac{1}{\beta}\sum_{\omega_n}\int \frac{d^3 p}{(2 \pi)^3} \frac{ e^{i P \cdot (X_1-X_2)}}{-P^2}.
\end{align}
Here $P\equiv (\omega_n, \vec{p})$, and: $P^2=-\omega_n^2-\vec{p}^2$. The Kubo-Martin-Schwinger relation states that the correlation functions are periodic in imaginary time with the period $\beta$. So, Matsubara frequencies are discrete and  are equal to $\omega_n=\frac{2 \pi n}{\beta}$. 

The propagator \eqref{green0MIN} is the building block for the perturbation theory in the Minkowski space-time. Consider the simplest example -- the one loop correction to the propagator. This correction  leads in particular to the generation of an effective mass:
\begin{gather}
\ \raisebox{-.4\height}{\includegraphics[scale = 0.9]{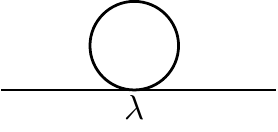} }\
\nonumber
= \\=
\label{oneloopMIN}
-\frac{\lambda}{2}\frac{1}{\beta}\sum_{\omega_n}\int \frac{d^3 p}{(2 \pi)^4} \frac{e^{i P\cdot (X_1-X_2) }}{(-P^2)^2} \int  \frac{d^3 k }{(2 \pi)^3} \frac{1}{|k|} \left(\frac{1}{2}+ \frac{1}{e^{\beta |k|}-1}\right).
\end{gather}
Here, the vector $\vec{k}$ represents the momentum inside the loop, while the vector $\vec{p}$ is the momentum in the external legs. The first term in the eq.\eqref{oneloopMIN} under the integral over $\vec{k}$ leads to a standard UV divergence and does not depend on the temperature. Such a state-independent divergence can be simply subtracted or absorbed into the UV mass counter-term. The only condition we imply at this point is that the physical mass is zero at $T=0$. So that the bare theory is conformal. 

We use the dimensional regularization through the paper. In this renormalization scheme, one can show that the first term (the vacuum term) under the integral over $\vec{k}$ in \eqref{oneloopMIN} gives zero. The second term under the integral in \eqref{oneloopMIN} is finite and is equal to zero at $T=0$. The subtraction procedure under consideration leads to the intuitively clear result. At $T=0$, the theory remains massless, while for $T>0$ a thermal mass does appear. 

Later in this section, we will solve the Dyson-Schwinger equation\footnote{In this case, it also coincides with a mass gap equation.}, which sums some type of diagrams build with the uses of \eqref{oneloopMIN}. Looking ahead, we write the expression for the resulting Debye mass here:
\begin{align}
\label{debyemassMIN}
\boxed{
 \quad   m_{thermal}^2 = \frac{ 1}{\beta^{2}}\bigg( \frac{\lambda}{24}-\frac{\lambda^{3/2}}{16\sqrt{6}\pi} \bigg)+O(\lambda ^2). \quad}
\end{align}
Remaining part of this chapter will be devoted to the physical meaning, analiticity properties of \eqref{debyemassMIN} as the function of $\lambda$, temperature dependence and to the physical manifestations of the thermal mass \eqref{debyemassMIN}.
\subsection{Physical meaning of the leading order}
\label{M1}
The expression \eqref{debyemassMIN} is obtained from the Dyson-Schwinger equation. It sums all diagrams of tadpole type. Of course, one can obtain the result in the leading order of $\lambda^1$ from a simpler reasoning with the clear physical meaning. Let us select some particle and see how it propagates and interacts with other particles in thermal bath, this process corresponds to the closed loop. Interaction with these particles is responsible for the thermal mass generation. On the one hand, the particles inside the loop in the bath also interact with the bath. From the other hand, such processes are suppressed by the higher order of $\lambda$. So the leading contribution to the thermal mass comes from the interaction with the thermal bath particles, which don't interact between themself. Indeed, consider the following subclass of the diagrams: 
\begin{figure}[H]
    \centering
    \includegraphics[scale=0.9]{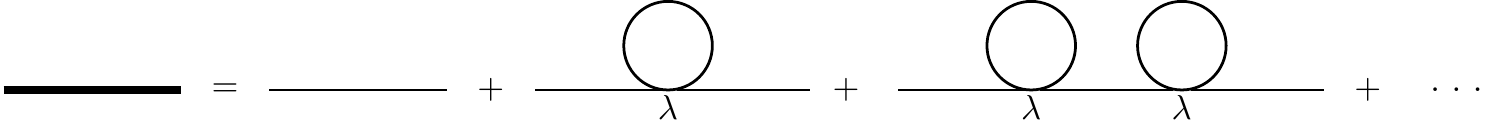}
    \caption{Direct sum of the tadpole diagrams in the Minkowski space-time. Each closed line is proportional to the factor $\beta^{-2}$.}
    \label{fig1}
\end{figure}
\noindent The straightforward summation of the diagrams from the Fig.\ref{fig1} gives the following expression to the corrected propagator at the first order:
\begin{gather}
    G_{\lambda^{1}}(X_1,X_2) = \frac{1}{\beta}\sum_{\omega_n}\int \frac{d^3 p}{(2 \pi)^3} \frac{e^{i P\cdot (X_1-X_2) }}{-P^2+m_{\lambda^{1}}^2},
\end{gather}
where the thermal mass at this order\footnote{We denote the order of calculations as $m^2=m^2_{\lambda^\alpha}+o(\lambda^\alpha)$.} is as follows:
\begin{align}
\label{DebyemassMIN1st}
m_{\lambda^{1}}^2=\frac{\lambda}{2} \int  \frac{d^3 k }{(2 \pi)^3} \frac{1}{|k|}  \frac{1}{e^{\beta |k|}-1} =\frac{\lambda}{24} \frac{1}{ \beta^2}.
\end{align}
As it should be, the last expression coincides with \eqref{debyemassMIN} in the first order.
\subsection{Higher order corrections and IR divergences}
\label{M2}
There is another moment that deserves attention. In the previous subsection we consider Debye mass in the leading order in the coupling constant $\lambda$ and obtain the expression \eqref{DebyemassMIN1st}. However, as one can see from eq.\eqref{debyemassMIN}, the next order correction is  $\lambda^{3/2}$, i.e. it is not an analytic function of $\lambda$. It is not a new phenomenon that solution of the Dyson-Schwinger equation appears to be a non-analytic function of the coupling constant. It means that a naive perturbation theory breaks down. This non-analyticity  has a clear physical origin in our situation -- there are IR divergence in the loops. For example, the following two loop diagram:
\begin{gather}
\raisebox{-.4\height}{\includegraphics[scale = 0.9]{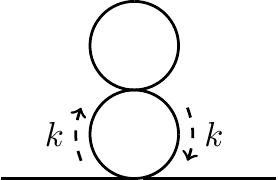} }
\propto \int \frac{ d^4 k}{k^4},
\end{gather}
has both IR and UV divergence. It is clear that UV divergence can be regularized as it is discussed before. But what about the IR one? The divergence arises because we consider massless bare theory. However, the proper loop  resummation (see Sec. \ref{32contributionMIN} for a detailed discussion in a similar situation in Rindler coordinates) or uses of the dressed propagators allows one to cancel the IR divergence. But, a remnant of the IR divergence manifests itself in the non-analyticity of the thermal mass \eqref{debyemassMIN} as a function of the coupling constant $\lambda$. 
\subsection{Dyson-Schwinger equation and the debye mass}
\label{M3}
In the remainder of this section, we consider the following Dyson-Schwinger equation:
\begin{figure}[H]
    \centering
    \includegraphics[scale=0.9]{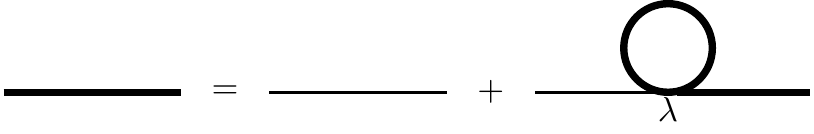}
    \caption{Dayson equation in the Minkowski space-time. Thick lines represent dressed or "exact" propagator. Thin lines are bare propagators.}
    \label{DysonEq}
\end{figure}
\noindent This equation can be solved straightforwardly by the following anzatz for the dressed propagator:

\begin{gather}
\label{modpropmin}
    G_{\lambda^{3/2}}(X_1,X_2) = \frac{1}{\beta}\sum_{\omega_n}\int \frac{d^3 p}{(2 \pi)^3} \frac{e^{i P\cdot (X_1-X_2) }}{-P^2+m^2}.
    \end{gather}
Where $m$ is an unknown thermal mass which should be found from the self consistent solution of the Dyson-Schwinger equation. Indeed, if one plugs \eqref{modpropmin} into the Dyson-Schwinger equation graphically shown on the Fig. \ref{DysonEq} one will get the following transcendental equation for $m$:
\begin{align}
\label{eqAlambda}
    m^2=\frac{\lambda}{16 \pi^2}\frac{m^2}{d-4}+\frac{\lambda m^2}{32\pi^2}\log\bigg(\frac{e^{-1+\gamma}m^2}{4\pi\mu^2}\bigg)+\frac{\lambda}{4 \pi^2}\int_0^\infty dp\frac{p^2 }{\sqrt{p^2+m^2}}\frac{1}{e^{\sqrt{p^2+m^2}}-1}.
\end{align}
Here $d-4$ is a parameter in the dimensional regularization, and $\mu$ is a dimensional parameter to keep $\lambda$ dimensionless: $\lambda\to\lambda\mu^{4-d}$. The first term on the right hand side (RHS) is divergent as $d\to4$. That is a problem, but what is worse is the state dependent divergence coming from the particle distribution/state dependent part of the contribution on RHS of \eqref{eqAlambda} . Because of that, the latter type of the divergence cannot be cancelled by the standard UV counter terms. However, this divergence is of the order $\lambda^2$. Hence, an improved perturbation theory is required to deal with the second order divergences. To consider $\lambda^2$ contribution one has to sum all the second order corrections: $a)$ a local thermal mass term with the the corrected $\lambda \phi^4$ vertex, $b)$ Nonlocal (sunset type) $\lambda^2$ loops: 
\begin{figure}[H]
    \centering
    \includegraphics[scale=0.9]{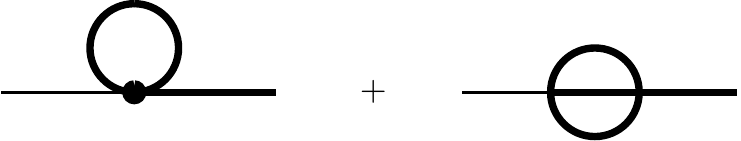}
    \caption{Loops of the order $\lambda^2$. Where thick dot is the corrected $\lambda \phi^4$ vertex.}
    \label{nonloc}
\end{figure}
This is a much more complex problem, especially in the Rindler space-time. For more details in Minkowski space-time see: \cite{ Parwani:1991gq,Drummond:1997cw,Peshier:1998rz, Blaizot:2003tw,  Kraemmer:2003gd, Andersen:2008bz, Arjun:2021gaf}. In the present article, we restrict ourselves to the order lower than $2$.

In all, as the divergence in eq. \eqref{eqAlambda} is of order $\lambda^2$ -- it can be dropped because it is beyond the one loop equation. The same situation with the second term on the RHS of eq. \eqref{eqAlambda}. Finally, using the eq. \eqref{eqAlambda} and the integral \eqref{J2beta} from the Appendix, one can find the expression for the thermal mass to the order of $\lambda^{3/2}$:
\begin{align}
\label{DebyemassMIN2st}
    m_{\lambda^{3/2}}^2 = \frac{1}{\beta^{2}}\bigg[ \frac{\lambda}{24}-\frac{\lambda^{3/2}}{16\sqrt{6}\pi} \bigg],
\end{align}
which leads to \eqref{debyemassMIN}.
\subsection{Stress–energy tensor}
\label{setMIN}
It is well-known that quantum expectation value of the stress-energy tensor in flat space is traceless for the free massless scalar field even at the finite temperature. However, in the previous section, we found that the spectrum does change due to loop corrections. In particular, we obtained the mass gap \eqref{DebyemassMIN2st}. But does the thermal mass lead to non-zero trace of the stress-energy tensor? On the one hand, the propagator \eqref{modpropmin} shows the massive spectrum. On the other hand, the classical expression of the stress-energy tensor following from the eq.\eqref{action} does not contain any mass term. 

To explain this controversy, we calculate the quantum expectation value of the stress energy tensor to find out how the loop corrections contribute to it. One can express the dressed expectation value of the stress energy tensor using the dressed propagator from the following definition:
\begin{align}
\label{genteimin}
    \langle \hat{T}_{\mu\nu}\rangle=\bigg(\frac{\partial}{\partial X_1^\mu}\frac{\partial}{\partial X_2^\nu}-\frac{1}{2}g_{\mu\nu}\frac{\partial}{\partial X_1^\alpha}\frac{\partial}{\partial X_{2\alpha}}\bigg)G_{\lambda^{3/2}}(X_1,X_2)\bigg|_{X_1=X_2} +\frac{\lambda}{8}g_{\mu\nu}\Big(G_{\lambda^{3/2}}(X_1,X_2)\Big)^2.
\end{align}
Where $G_{\lambda^{3/2}}(X_1,X_2)$ is the resummed propagator that we have found above. Simple substitution gives:  
\begin{align}
\langle \hat{T}_{\mu \nu}\rangle= \frac{1}{\beta}\sum_{\omega_n}\int \frac{d^3 p}{(2 \pi)^3}  \frac{P_\mu P_\nu -\frac{1}{2} g_{\mu\nu} P^2}{-P^2+m_{\lambda^{3/2}}^2}+\frac{\lambda}{8} g_{\mu\nu} \left(\frac{1}{\beta}\sum_{\omega_n}\int \frac{d^3 p}{(2 \pi)^3} \frac{1}{-P^2+m_{\lambda^{3/2}}^2}\right)^2.
\end{align}
The latter expression has an UV divergences. As it is discussed above we restrict ourselves to the $\lambda$ and $\lambda^{3/2}$ corrections and drop off $\lambda^2$ and higher powers. After dimensional regularization the divergences are proportional to $m^4/\epsilon$ or to $\lambda m^2/\epsilon$. Because $m^2\sim \lambda$, they are of the second order in $\lambda^2$(see, for example, \cite{BarrosoMancha:2020fay}) and are beyond the order at which we work in this paper. Technical details of the dimensional regularization are given in the Appendix \ref{AppDressedPropReg}. And the regularized expectation value of the stress energy tensor at the order that we work in is as follows:
\begin{align}
\label{setmin}
\boxed{\ \ 
\langle \hat{T}_{\nu}^\mu\rangle^{reg}=   T^4\bigg[ \frac{\pi^2}{90}-\frac{1}{48}\frac{\lambda}{ 4! }+\frac{1}{12 \pi} \left(\frac{\lambda}{4!}\right)^{3/2}\bigg] \begin{pmatrix}
3 & 0 & 0 & 0\\
0 & -1 & 0 & 0\\
0 & 0 & -1 & 0\\
0 & 0 & 0 & -1\\
\end{pmatrix}+O(\lambda^2). \ \ }
\end{align}
The obtained result is consistent with \cite{Nachbagauer:1995wn}. As one can see, this tensor is traceless. So, the thermal mass generates the mass gap in the spectrum, but does not generate the trace of the quantum average of the stress energy tensor and the theory itself remains conformal. In that space-time stress energy tensor should remain traceless even on the quantum level in accordance with conformal anomaly. 

\section{Thermal states in the Rindler space-time}
There is an open question about whether the thermal equilibrium states with the planckian distribution for the exact mode in the Rindler space-time with an arbitrary inverse temperature $\beta$ have standard thermal properties \cite{Akhmedov:2020qxd, Akhmedov:2020ryq, Anempodistov:2020oki, Akhmedov:2021cwh, Bazarov:2021rrb}. As an attempt to address this issue we calculate the Debye mass for an arbitrary temperature and compare the result with the answer obtained for the thermal state in the Minkowski space-time. So, in this section, we repeat in the Rindler coordinates the calculation from the Sec.\ref{debmassMINsec}.
\subsection{Quantization} 
\label{gensecRindler}
Let us start with a brief overview of the definitions and basic properties of the thermal propagator in the Rindler coordinates (for more details see: \cite{Unruh:1983ms, Kay:1985zs, Kay:1985yw, Moschella:2009ub, Bazarov:2021rrb}). The Rindler coordinates can be obtained from the Minkowskian ones by the following transformation: $t=e^\xi \sinh \eta, \ \  x=e^\xi \cosh \eta $ (the proper acceleration is set to one). These coordinates cover only quarter $x>|t|$ of the entire Minkowski space-time. As a result the induced Rindler metric has the following form:
\begin{align}
\label{metricRin}
    ds^2=e^{2\xi}(d\eta^2-d\xi^2)-d\vec{z}^2.
\end{align}
Where $\vec{z}$ is a two dimensional vector in the transverse directions to the $\eta-\xi$ plane, and $\eta$ is the time-like coordinate.  Let us introduce the following notation: $X=(\xi,\eta,\vec{z})$. Another object, which is interesting for us, is the squared geodesic distance $\sigma$ between two points $X_1$ and $X_2$ in the metric \eqref{metricRin}:
\begin{align}
\label{geodesic}
    \sigma\left(X_1, X_2\right) =2e^{\xi_1+\xi}\cosh(\eta_2-\eta_1)-e^{2\xi_1}-e^{2\xi_2}-|\vec{z}_2-\vec{z}_1|^2.
\end{align}

Before we go further, let us say a few words about the geometry under consideration. The Rindler wedge is a globally hyperbolic manifold, but it is geodesically incomplete. Moreover Cauchy surfaces in the Rindler and Minkowski coordinates are different. Rindler Cauchy surfaces cannot be used for the initial value problem in the entire Minkowski space-time. As a result the set of modes in Rindler space-time is not a complete set for the entire Minkowski space-time. On the other hand, the set of modes in the the Minkowski wedge is overcomplete for the Rindler space-time. Hence, the Minkowskian modes are not correctly normalized in Rindler wedge.

The metric \eqref{metricRin} has a time-like killing vector $\partial_\eta$, which corresponds to the time translational symmetry. This fact allows us to introduce the thermal equilibrium state with an arbitrary temperature. Note, however, that in the limit $\xi\to-\infty$ the metric \eqref{metricRin} degenerates, and in this case we have a coordinate singularity where time-like killing vector becomes light-like.

We consider the free scalar field (the action \eqref{action} with $\lambda=0$). The corresponding quantum field operator has the following form:
\begin{align}
\label{operatorRindler}
    \hat{\varphi}(X) =\int_{-\infty}^{+\infty} \frac{d^2k}{2\pi} \int_{0}^{+\infty} \frac{d \omega}{\pi}\sqrt{\sinh\pi \omega}\bigg[e^{-i\omega \eta+i\vec{k}\vec{z}}\hat{a}_{\omega,\vec{k}}^{}  +e^{i\omega \eta- i\vec{k}\vec{z}}\hat{a}_{\omega,\vec{k}}^\dagger\bigg]K_{i\omega}\big(k e^{\xi}\big),
\end{align}
where $K_{i \omega}\big(k e^{\xi}\big) $ is the MacDonald function and the creation and annihilation operators obey the standard commutation relations:
\begin{align}
    [\hat{a}_{\omega,\vec{k}}^{},\hat{a}_{\omega',\vec{k'}}^\dagger]=\delta(\w-\w')\delta^2(\vec{k}-\vec{k'}); \qquad [\hat{a}_{\omega,\vec{k}}^\dagger,\hat{a}_{\omega',\vec{k'}}^\dagger]=0; \qquad  [\hat{a}_{\omega,\vec{k}}^{},\hat{a}_{\omega',\vec{k'}}^{}]=0.
\end{align}
The thermal equilibrium state with the inverse temperature $\beta\equiv T^{-1}$ is represented by the following density operator: 
\begin{align}
\label{densmat}
\hat{\rho}=\frac{e^{-\beta \hat{:H:}}}{\text{Tr} e^{-\beta \hat{:H:}}}  , \quad \text{where} \quad \hat{:H:}= \int_0^\infty d\w \w  \int d^2k \hat{a}^\dagger_{\w,\vec{k}}\hat{a}^{}_{\w,\vec{k}}.
\end{align}
And the expectation value of an operator is defined as $  \langle\hat{O}\rangle_\beta\equiv \text{Tr}\hat{\rho}\hat{O}$.  Then one can directly compute the Wightman function:
\begin{gather}
\label{Wbeta}
     W_\beta(X_1,X_2)=\langle\varphi(X_1)\varphi(X_2)\rangle_\beta
     =\\=
     \nonumber
     \int_{0}^{+\infty} \frac{d^2k d\omega}{(2\pi)^2\pi^2} \sinh(\pi \omega) \frac{1}{e^{\beta \omega}-1} e^{i \omega (\eta_1-\eta_2)}e^{i\vec{k} (\vec{z}_1-\vec{z}_2)} K_{i\omega}\big(k e^{\xi_1}\big)K_{i\omega}\big(k e^{\xi_2}\big)
     +\\+
     \nonumber
     \int_{0}^{+\infty} \frac{d^2k d\omega}{(2\pi)^2\pi^2}\sinh(\pi \omega) \left[1+ \frac{1}{e^{\beta \omega}-1}\right]   e^{-i \omega (\eta_1-\eta_2)}e^{i\vec{k} (\vec{z}_1-\vec{z}_2)} K_{i\omega}\big(k e^{\xi_1}\big)K_{i\omega}\big(ke^{\xi_2}\big).
\end{gather}
There is a certain temperature -- so called Unruh temperature ( $T_U^{-1}=\beta_U=2\pi$, where we set acceleration to one), that is experienced by a uniformly accelerated observer in the Minkowski vacuum. In other words the Minkowski (Poincare invariant) vacuum is the thermal state with the Unruh temperature from the point of view of an accelerated observer. Using direct calculation, one can verify that the vacuum two point function in the Minkowski space-time equals to the thermal two point function with the Unruh temperature in the Rindler space-time \cite{Unruh:1976db}. The quick way to see this is as follows. We know that vacuum two point function in the Minkowski space-time depends on the geodesic distance $\sqrt{\sigma}$. Hence, in terms of Rindler coordinates it is periodic in the complex time plane: 
\begin{align}
\label{2pisym}
    \eta_2-\eta_1 \to  \eta_2-\eta_1 + 2\pi i.
\end{align}Then the KMS condition tells us that the density matrix for fields in Rindler space-time is \eqref{densmat}  with $\beta=\beta_U$. 
 Furthermore, in \cite{Higuchi:2020swc,Akhmedov:2022uug} (see also \cite{Higuchi:2010xt,Akhmedov:2021agm} for the related topic) the agreement between interacting field theory in the Unruh state in the Rindler space-time and interacting field theory in the Poincare invariant vacuum in the Minkowski space-time is showed. Hence, for these states even loop corrected propagators can be analytically continued from the Rindler quarter of the entire Minkowski space-time. Indeed, all the results of our paper are in agreement of that statement.

It is also worth saying a few words about the massive scalar field in the Rindler space-time. The field operator has the following form (for more details see \cite{birrell_davies_1982,Akhmedov:2020ryq}):
\begin{align}
\label{massivefieldRin}
    \hat{\varphi}(X) =\int_{-\infty}^{+\infty} \frac{d^2k}{2\pi} \int_{0}^{+\infty} \frac{d \omega}{\pi}\sqrt{\sinh\pi \omega}\bigg[e^{-i\omega \eta+i\vec{k}\vec{z}}\hat{b}_{\omega,\vec{k}}^{}  +e^{i\omega \eta- i\vec{k}\vec{z}}\hat{b}_{\omega,\vec{k}}^\dagger\bigg]K_{i\omega}\big(\sqrt{m^2+k^2} e^{\xi}\big).
\end{align}
We use the symbol $\hat{b}$ to denote the creation and anihilation operators and show the differences between the massless and massive cases. The thermal Wightman function in this case is as follows:
\begin{gather}
\label{masGreenfunction}
     W_\beta(X_1,X_2)
     =\\=
     \nonumber
     \int_{0}^{+\infty} \frac{d^2k d\omega}{(2\pi)^2\pi^2} \sinh(\pi \omega) \frac{1}{e^{\beta \omega}-1} e^{i \omega (\eta_1-\eta_2)}e^{i\vec{k} (\vec{z}_1-\vec{z}_2)} K_{i\omega}\big(\sqrt{m^2+k^2} e^{\xi_1}\big)K_{i\omega}\big(\sqrt{m^2+k^2} e^{\xi_2}\big)
     +\\+
     \nonumber
     \int_{0}^{+\infty} \frac{d^2k d\omega}{(2\pi)^2\pi^2}\sinh(\pi \omega) \left[1+ \frac{1}{e^{\beta \omega}-1}\right]   e^{-i \omega (\eta_1-\eta_2)}e^{i\vec{k} (\vec{z}_1-\vec{z}_2)} K_{i\omega}\big(\sqrt{m^2+k^2} e^{\xi_1}\big)K_{i\omega}\big(\sqrt{m^2+k^2} e^{\xi_2}\big).
\end{gather}
The key point here, is that the spectrum of the field theory under consideration starts from zero (see for example \cite{Bazarov:2021rrb}), i.e. even for the massive theory there is no mass gap. Indeed, the Hamiltonian operator is given by the following expression:
\begin{align}
\label{hamiltonianRin}
    \hat{:H:}= \int_0^\infty d\w \w  \int d^2 k  \hat{b}^\dagger_{\w,\vec{k}}\hat{b}^{}_{\w,\vec{k}}.
\end{align}
As one can see, that it is similar to the massless case. The reason for that is the presence of the horizon in the vicinity of which the mass term in the action can be neglected.

Using the thermal Wightman function, one can construct the thermal Feynman propagator as
follows: 
\begin{align}
G_{\lambda^0}(X_1,X_2)=\theta(\eta_1-\eta_2) W_\beta(X_1,X_2)+\theta(\eta_2-\eta_1) W_\beta(X_2,X_1),
\end{align}
which after the analytical continuation to the Euclidean signature has the following form:
\begin{multline}
\label{green0RIn}
    G_{\lambda^{0}}(X_1,X_2)= \\=\frac{1}{\beta}\sum_{\omega_n}\int \frac{d^2 \vec{k}}{(2 \pi)^2} \int_0^{\infty} \frac{ d \omega}{\pi^2} \frac{2\omega\sinh\pi \omega}{\w_n^2+\omega^2}e^{-i\w_n (\eta_2-\eta_1)}e^{i\vec{k}(\vec{z}_2-\vec{z}_1)}K_{i \omega}\big(ke^{\xi_1}\big)K_{i\omega}\big(ke^{\xi_2}\big),
\end{multline}
where: $\omega_n=\frac{2 \pi n}{\beta}$ are the Matsubara frequencies. To obtain the last expression we use the following relation:
\begin{align}
\frac{1}{\beta} \sum_{\omega_n} f(\omega_n)= \int_{-\infty}^\infty \frac{d w}{2\pi} f(w)
+
\int_{-\infty-i 0^+}^{\infty-i 0^+} \frac{d w}{2\pi}\left[ f(w)+ f(-w)\right]  \frac{1}{e^{  i\beta w}-1}.
\end{align}

\subsection{Pertrubative one loop thermal mass in Rindler space-time}
\label{massRindler}
In this section we consider one loop correction to the propagator in the Rindler space-time. One loop correction contains the Feynman propagator \eqref{green0RIn} at coincident points which, together with the two external propagators, must be integrated over the space-time volume. Therefore, we start with the consederation of the propagator at coincident points:
\begin{align}
\label{propxxrin}
G_{\lambda^{0}}(X,X) = \frac{1}{\pi^2}\int\frac{ d^2k}{(2\pi)^2}\int_0^{+\infty}d\w K_{i\w}\big(ke^{\xi}\big)^2\sinh\pi\w\bigg[1+\frac{2}{e^{\beta \omega}-1}\bigg],
\end{align}
where we perform the summation over the Matsubara frequencies. The first term under the integral of this equation leads to the UV divergence due to the zero-point fluctuations. In the Minkowski space-time \eqref{oneloopMIN} the first (vacuum) part vanishes after the dimensional regularization, while the second (thermal) part is finite. We have seen this in the previous section and explain in the Appendix. This is an expected result -- there is no contribution to the thermal mass at zero temperature.

As it is stated in Sec.\ref{gensecRindler}, in the Rindler space-time the state with inverse temperature $\beta=2\pi$ corresponds to the Minkowski vacuum. Hence, one should expect that the vacuum part of the propagator  \eqref{propxxrin} leads to a finite contribution after the dimensional regularization, that should somehow cancel thermal contribution for the case $\beta=2\pi$. Indeed, after the dimensional regularization (see Appendix \ref{AppBarePropReg}) the propagator at coincident points is:
\begin{align}
\label{regK}
\ \ \raisebox{-.4\height}{\includegraphics[scale = 0.9]{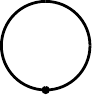}} \ \ =\frac{e^{-2\xi}}{12}\Bigg(\frac{1}{\beta^2}-\frac{1}{(2\pi)^2}\Bigg).
\end{align}
Then, for the temperature $\beta=2\pi$ the thermal mass does vanish. This is consistent with the fact that for the Poincaré invariant state (Minkowski vacuum), the thermal mass is zero. Similar result is obtained in the static de Sitter space-time \cite{Popov:2017xut}. 

Hence, using \eqref{regK}, it is straightforward to find the one loop correction: 
\begin{gather}
\raisebox{-.4\height}{\includegraphics[scale = 0.9]{graph_debye_2.pdf} }
= \\=
\nonumber
\frac{1}{\beta}\sum_{\omega_n}\int \frac{d^2 \vec{k}}{(2 \pi)^2} \int_0^{\infty} \frac{ d \omega}{\pi^2} \frac{2\omega\sinh\pi \omega}{\w_n^2+\omega^2} e^{-i\w_n (\eta_2-\eta_1)}e^{i\vec{k}(\vec{z}_2-\vec{z}_1)}K_{i \omega}\big(ke^{\xi_1}\big)K_{i\omega}\big(ke^{\xi_2}\big) \left[-\frac{\frac{\lambda}{24}\big(\frac{1}{\beta^2}-\frac{1}{(2\pi)^2}\big)}{\w_n^2+\omega^2}\right] . 
\end{gather}
The latter is very similar to the expression for the bare propagator \eqref{green0RIn}, but with an additional factor in the square brackets. It is probably worth stressing that such a simple form of the loop correction does not appears in the massive field theory. For more details see Appendix \ref{AppBarePropReg}. 

Furthermore, one can perform the resummation of the subclass of diagrams from the Fig.\ref{fig1}. Trivial summation gives the following corrected propagator:
\begin{gather}
 G_{\lambda^1} (X_1,X_2) =
    \frac{1}{\beta}\sum_{\omega_n}\int \frac{d^2 \vec{k}}{(2 \pi)^2} \int_0^{\infty} \frac{ d \omega}{\pi^2} \frac{2\omega\sinh\pi \omega}{\w_n^2+\omega^2+M_{\lambda^1}^2}e^{-i\w_n (\eta_2-\eta_1)}e^{i\vec{k}(\vec{z}_2-\vec{z}_1)}K_{i \omega}\big(ke^{\xi_1}\big)K_{i\omega}\big(ke^{\xi_2}\big),
\end{gather}
where $M_{\lambda^1}^2$ is a squared thermal mass in the first order in $\lambda$:
\begin{align}
\label{mass12}
M_{\lambda^1}^2=\frac{\lambda}{24}\left(\frac{1}{\beta^2}-\frac{1}{(2 \pi)^2}\right).
\end{align}
Evaluation of the sum over the Matsubara frequencies gives: 
\begin{gather}
    G_{\lambda^1}(X_1,X_2)=  \int \frac{d^2 \vec{k}}{(2 \pi)^2} \int_0^{\infty} \frac{ d \omega}{\pi^2} \frac{\omega\sinh\pi \omega}{E_\omega}e^{i\vec{k}(\vec{z}_2-\vec{z}_1)}K_{i \omega}\big(ke^{\xi_1}\big)K_{i\omega}\big(k e^{\xi_2}\big)
    \times\\ \times
    \nonumber
  \left[e^{ -E_\omega (\eta_2-\eta_1)}\left(1+ \frac{1}{e^{\beta E_\omega }-1}\right)+e^{ E_\omega (\eta_2-\eta_1)}\frac{1}{e^{\beta E_\omega }-1}\right],
\end{gather}
where:
\begin{align}
\label{E}
E_\omega=\sqrt{\omega^2+M_{\lambda^1}^2}.
\end{align}
Thus, taking into account loop corrections leads to the appearance of the gap in the spectrum.

Let us stress at this point that the key difference between the revelation of the thermal masses in Minkowski and Rindler coordinates follows from the fact that if one deals with the exact modes for generic $\beta$ one deals with different states. Compare the contributions of the thermal and physical masses to the propagators. The thermal mass $M_{thermal}$ is coming from the loop corrections. The physical mass $M_{phys}$ is coming from the additional term  $M_{phys}^2\varphi^2/2$ in the action (recall, that the massive Wightman function in the Rindler space-time is given by \eqref{masGreenfunction}). Then, the effect from these masses on the bare propagators \eqref{green0MIN} and \eqref{green0RIn} can be schematically shown in the table:
\begingroup
\renewcommand*{\arraystretch}{2.2}
\begin{center}
\begin{tabular}[3]{ |c|c|c| } 
 \hline
  & Physical mass $M_{phys}$ & Thermal mass $M_{thermal}$  \\ 
  \hline
 Minkowski & $\frac{1}{-P^2} \to \frac{1}{-P^2+M_{phys}^2}$ & $\frac{1}{-P^2} \to \frac{1}{-P^2+M_{thermal}^2} $\\ 
  \hline
 Rindler & $K_{i\omega}\big(ke^{\xi_2}\big)\to K_{i\omega}\big(\sqrt{k^2+M_{phys}^2}e^{\xi_2}\big)$ & $\frac{1}{\w_n^2+\omega^2}\to\frac{1}{\w_n^2+\omega^2+M_{thermal}^2}$ \\ 
 \hline
\end{tabular}
\end{center}
\endgroup
It can be seen, that in the Minkowski space-time physical and thermal masses have the same effect on the propagator, while in the Rindler coordinates they don't. For example, accordingly to \eqref{hamiltonianRin} the spectrum of the massive field theory in the Rindler coordinates starts from zero, but this property does not hold in the presence of the thermal mass due to \eqref{E}.

It should also be noted that if the temperature is less than the Unruh one, then the mass squared in \eqref{mass12} appears to be negative, and therefore $E_\w$ becomes complex. So, the theory becomes unstable due to the presence of the tachyonic mass. Similar property for the Rindler vacuum $(\beta=\infty)$ was discussed in \cite{Stephens:1986ie}.

\subsection{Dyson-Schwinger equation in the Rindler space-time}
\label{DS Rind}
Now let us consider the Dyson-Schwinger equation for the dressed propagator which is shown on the Fig.(\ref{DysonEq}). It is very similar to the equation in the Minkowski space-time. The key difficulty now is the choice of the regularization procedure. As one can see, the regularization of the bare propagator \eqref{regK} at the coincident points is equivalent to the subtraction of the propagator with the Unruh temperature. Naively, one would expect that for the Dyson-Schwinger equation from the Fig.(\ref{DysonEq}) the regularization is reduced simply to the subtraction of the contribution with the Unruh temperature, i.e. subtraction of the contribution in the Minkowski vacuum. As we will see below, however, dimensional regularization beyond one loop leads to the fact that the Debye mass and the stress energy tensor are not simply functions of the difference of the squared inverse temperature as in \eqref{regK}. 

The Dyson-Schwinger equation can be solved straightforwardly by the following anzatz for the dressed propagator:

\begin{gather}
\label{l32}
    G_{\lambda^{3/2}}(X_1,X_2) =\\=
    \nonumber
    \frac{1}{\beta}\sum_{\omega_n}\int \frac{d^2 \vec{k}}{(2 \pi)^2} \int_0^{\infty} \frac{ d \omega}{\pi^2} \frac{2\omega\sinh\pi \omega}{\w_n^2+\omega^2+M^2}e^{-i\w_n (\eta_2-\eta_1)}e^{i\vec{k}(\vec{z}_2-\vec{z}_1)}K_{i \omega}\big(ke^{\xi_1}\big)K_{i\omega}\big(ke^{\xi_2}\big),
    \end{gather}where $M$ is an unknown thermal mass which should be found from the Dyson-Schwinger equation. 
 
After the dimensional regularization (see Appendix \ref{AppBarePropReg}) the dressed propagator at coincident points acquires the form:
\begin{align}
\label{regKdres}
\ \ \raisebox{-.4\height}{\includegraphics[scale = 0.9]{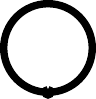}} \ \ =
 e^{-2 \xi } \left[\frac{1}{12 }\left(\frac{1}{\beta^2}-\frac{1}{4 \pi ^2}\right)-\frac{M}{4  \pi\beta }\right]+O(M^2)
\end{align}

If one plugs \eqref{regKdres} and \eqref{l32} into the Dyson-Schwinger equation graphically shown on the Fig.(\ref{DysonEq}) one will get the following equation:
\begin{gather}
\label{gapRin}
    M^2=\frac{\lambda}{24 }\left(\frac{1}{\beta^2}-\frac{1}{(2 \pi)^2}\right)-\frac{\lambda}{2}\frac{M}{4  \pi\beta }+O(\lambda^2).
\end{gather}
Solving \eqref{gapRin} for $M$ one can find the next order contribution to the thermal mass. As in the Minkowski space-time naive perturbation series breaks down. As a result the thermal mass becomes a non-analytic function of $\lambda$: 
\begin{align}
\boxed{ \ \ 
\label{masdeb3/2}
M_{\lambda^{3/2}}^2=\frac{\lambda}{24}\left(\frac{1}{\beta^2}-\frac{1}{(2 \pi)^2}\right)- \frac{\lambda^{3/2}}{16 \sqrt{6} \pi} \frac{1}{\beta } \sqrt{\frac{1}{\beta^2}-\frac{1}{(2 \pi)^2}}+O(\lambda^2). \ \ }
\end{align}
It should also be noted that if we take the high temperature limit $\beta\to0$, then the thermal mass becomes exactly the same as in the Minkowski space-time \eqref{DebyemassMIN2st}. 
\subsection{Another way to obtain $\sim \lambda^{3/2}$ contribution in $M^2$ }
\label{32contributionMIN}
Expression \eqref{gapRin} follows from the Dyson-Schwinger equation. To obtain it we have used dimensional regularization to regulate the dressed propagator in the loops. Let as show that the non-analytical term in \eqref{gapRin} follows from the direct summation of diagrams. Consider the following subclass of the loop corrections to the propagator at the coincident points with at least one loop with bare propagator:
\begin{gather}
 \sum_{l=1}^\infty
\ \ \ \raisebox{-.4\height}{\includegraphics[scale = 0.8]{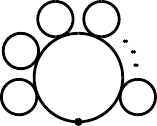} }\
\nonumber
= \\=
\frac{1}{\beta}\sum_{\omega_n}\int \frac{d^2 \vec{k}}{(2 \pi)^2} \int_0^{\infty} \frac{ d \omega}{\pi^2} \frac{2\omega\sinh\pi \omega}{\w_n^2+\omega^2} K_{i \omega}\big(ke^{\xi_1}\big)K_{i\omega}\big(ke^{\xi_2}\big)
\sum_{l=1}^{\infty}\left[-\frac{\frac{\lambda}{24}\big(\frac{1}{\beta^2}-\frac{1}{(2\pi)^2}\big)}{\w_k^2+\omega^2}\right]^l.
\end{gather}By adding and subtracting the term without loops, that we have considered separately in \eqref{regK}, one obtains the following expression:
\begin{gather}
 \sum_{l=0}^\infty
\ \ \ \raisebox{-.4\height}{\includegraphics[scale = 0.8]{graph_debye_9.pdf} }\ \ - \ \ \raisebox{-.4\height}{\includegraphics[scale = 0.9]{graph_debye_10.pdf}} \ \  = 
\nonumber
\frac{1}{\beta}\sum_{\omega_n}\int \frac{d^2 \vec{k}}{(2 \pi)^2} \int_0^{\infty} \frac{ d \omega}{\pi^2} 2\omega\sinh\pi \omega K_{i \omega}\big(ke^{\xi_1}\big)K_{i\omega}\big(ke^{\xi_2}\big)  
\times\\ \times
\label{wn}
\left(\frac{1}{\w_n^2+\omega^2+\frac{\lambda}{24}\big(\frac{1}{\beta^2}-\frac{1}{(2\pi)^2}\big) }-\frac{1}{\w_n^2+\omega^2}\right). 
\end{gather}
Each of the terms with $\w_n>0$ in \eqref{wn} contributes at the order $\lambda$, since they are IR finite, but we are interested in terms  that are proportional to $\lambda^\frac{1}{2}$ . The term with $\w_n=0$ is IR divergent for any number of the loops. But the divergences cancel only in the sum of all the terms under consideration. Hence, the contribution of the order  $\lambda^{\frac{1}{2}}$ comes from the modes with zero Matsubara frequency. If we single out only those contributions, we obtain:
\begin{gather}
\nonumber
\frac{1}{\beta}\int \frac{d^2 \vec{k}}{(2 \pi)^2} \int_0^{\infty} \frac{ d \omega}{\pi^2} 2\omega\sinh\pi \omega K_{i \omega}\big(ke^{\xi_1}\big)K_{i\omega}\big(ke^{\xi_2}\big)  
\left(\frac{1}{\omega^2+\frac{\lambda}{24}\big(\frac{1}{\beta^2}-\frac{1}{(2\pi)^2}\big) }-\frac{1}{\omega^2}\right) 
=\\=
-e^{-2 \xi} \frac{\lambda^{\frac{1}{2}}}{8 \sqrt{6} \pi \beta} \sqrt{\frac{1}{\beta^2}-\frac{1}{(2 \pi)^2}}.
\end{gather}
Thus, by the direct computation, we obtain the expression for the resumed propagator at the coincident points up to $\lambda^{1/2}$ order : 
\begin{gather}
G_{\lambda^1}(X,X)=\raisebox{-.4\height}{\includegraphics[scale = 0.9]{graph_debye_10.pdf}}+\sum_{l=1}^\infty
\  \raisebox{-.4\height}{\includegraphics[scale = 0.8]{graph_debye_9.pdf}} =e^{-2 \xi} \left[\frac{1}{12}\Bigg(\frac{1}{\beta^2}-\frac{1}{(2\pi)^2}\Bigg) - \frac{\lambda^{\frac{1}{2}}}{8 \sqrt{6} \pi \beta} \sqrt{\frac{1}{\beta^2}-\frac{1}{(2 \pi)^2}} \right]+O(\lambda).
\end{gather}
That is exactly the same expression as in \eqref{masdeb3/2}.

\subsection{Stress–energy tensor in the Rindler coordinates }
\label{SecsetRin}

Here we calculate the quantum expectation value of the stress-energy tensor over the thermal states under consideration. The method is similar to the one used in Minkowski coordinates. But in the Rindler coordinates it is principally important to take into account the $\XI$ term from \eqref{action}. This term does not contribute to the equations for motion because $R=0$. But the classical expression of the stress energy tensor defined as the variation of the action \eqref{action} with respect to the metric has the following form:
\begin{align}
\label{Txi}
T_{\mu\nu}=\partial_\mu\varphi\partial_\nu\varphi-g_{\mu\nu}\Big(\frac{1}{2}\partial_{\alpha}\varphi\partial^{\alpha}\varphi-\frac{\lambda}{4!}\varphi^4\Big)+\XI \Big(g_{\mu\nu}\nabla_\alpha\nabla^\alpha\varphi^2-\nabla_\mu\nabla_\nu \varphi^2\Big).
\end{align}
 The $\XI $ term gives a non-zero contribution because, roughly speaking, $\langle \varphi^2\rangle$ depends on the $\xi$ Rindler coordinate. In this article we focus on the conformal field theory case, where $\XI=\frac{1}{6}$. Moreover it turnes out that the regularization procedure in this case is simpler than for the other values of $\XI$. To clarify this point, consider the quantum average of the stress-energy tensor of a non-interacting massless field \cite{Bazarov:2021rrb}:
\begin{multline*}
    \langle:\hat{T}^{\mu}_{\nu}:\rangle_{\beta=\frac{1}{T}}\approx\frac{1}{480\pi^2}e^{-4\xi}
\Big((2\pi T)^4-1\Big)\begin{pmatrix}
1 & 0 & 0 & 0\\
0 & -\frac13 & 0 & 0\\
0 & 0 & -\frac13 & 0\\
0 & 0 & 0 & -\frac13\\
\end{pmatrix}+\\+\frac{1}{48\pi^2}\Big(1-6\XI\Big)e^{-4\xi}
\Big((2\pi T)^2-1\Big)\begin{pmatrix}
1 & 0 & 0 & 0\\
0 & -\frac13 & 0 & 0\\
0 & 0 & \frac23 & 0\\
0 & 0 & 0 & \frac23\\
\end{pmatrix}, \quad \text{as} \quad \xi\to -\infty.
\end{multline*}
It contains two different tensor strictures. Roughly speaking, the first one is proportional to $T^4$, and the second one is proportional to $T^2T_c^2$ (here $T_c$ is the canonical temperature). In the dimensional regularization, loop corrections to the first term diverge as $M^4/\epsilon$, while corrections to the second term behave as $M^2T_c^2/\epsilon$. This can be understood on the dimensional grounds. Note that $M^4$ is proportional to the second order $\lambda^2$, and it is beyond our consideration. However, $M^2/\epsilon$ is the first order term in $\lambda$ and should be carefully treated. Fortunately in the case $\XI=\frac{1}{6}$ the coefficient in front of the term $T^2T_c^2$ is zero \cite{Moretti:1997qn,PhysRevD.55.3552} and the regularization procedure becomes easier. The $\XI\ne\frac{1}{6}$ case will be considered separately, since it requires a renormalization of the constant $\XI$.

We calculate the quantum average of the operator \eqref{Txi} using the dressed propagator obtained in the previous section:
\begin{gather}
    \langle \hat{T}_{\mu\nu}\rangle=
    \bigg(\frac{\partial}{\partial X_1^\mu}\frac{\partial}{\partial X_2^\nu}-\frac{1}{2}g_{\mu\nu}\frac{\partial}{\partial X_1^\alpha}\frac{\partial}{\partial X_{2\alpha}}\bigg)G_{\lambda^{3/2}}(X_1,X_2)\bigg|_{X_1=X_2} 
    +\\+
    \nonumber
    \frac{\lambda}{8}g_{\mu\nu}\Big(G_{\lambda^{3/2}}(X_1,X_2)\Big)^2\bigg|_{X_1=X_2} +\frac{1}{6} \Big(g_{\mu\nu}\nabla_\alpha\nabla^\alpha-\nabla_\mu\nabla_\nu \Big)G_{\lambda^{3/2}}(X,X).
\end{gather}

The dressed propagator contains both the divergent vacuum term and the thermal term. The first one should be properly regularized (see Appendix \ref{AppDressedPropReg} for more details). Using regularized expression for dressed propagator \eqref{der0} and its derivatives with respect to coordinates \eqref{derz}, \eqref{derxi} and \eqref{dereta}, it is straightforward to write the quantum average of the stress energy tensor :
\begin{empheq}[box=\widefbox]{multline}
\label{setRin}
\langle T^{\mu}_\nu\rangle^{reg}=
e^{-4\xi}\bigg[\frac{\pi^2}{90}\Big(\frac{1}{\beta^4}-\frac{1}{(2\pi)^4}\Big)-\frac{1}{48}\frac{\lambda}{4!}\Big(\frac{1}{\beta^2}-\frac{1}{(2\pi)^2}\Big)^2+\\+\frac{1}{12 \pi}\left(\frac{\lambda}{4!}\right)^{3/2} \frac{1}{\beta}\Big(\frac{1}{\beta^2}-\frac{1}{(2\pi)^2}\Big)^{\frac{3}{2}}\bigg] \begin{pmatrix}
3 & 0 & 0 & 0\\
0 & -1 & 0 & 0\\
0 & 0 & -1 & 0\\
0 & 0 & 0 & -1\\
\end{pmatrix}+O(\lambda^2).
\end{empheq}
For the case $\lambda=0$, the obtained result agrees with \cite{PhysRevD.51.2770,PhysRevD.55.3552,Moretti:1997qn,Bazarov:2021rrb}. It is convenient to rewrite \eqref{setRin} in terms of the thermal mass:
\begin{align}
\langle T^{\mu}_\nu\rangle^{reg}=e^{-4\xi}\bigg[\frac{\pi^2}{90}\Big(\frac{1}{\beta^4}-\frac{1}{(2\pi)^4}\Big)-\frac{M_{\lambda^{3/2}}^2}{48}\Big(\frac{1}{\beta^2}+\frac{1}{(2\pi)^2}\Big)+\frac{M_{\lambda^{3/2}}^3}{48\beta\pi}\bigg] \begin{pmatrix}
3 & 0 & 0 & 0\\
0 & -1 & 0 & 0\\
0 & 0 & -1 & 0\\
0 & 0 & 0 & -1\\
\end{pmatrix}+O(\lambda^2).
\end{align}
The obtained stress energy tensor has several important properties. First, it remains traceless even for the self-interacting theory. Second, it diverges near the horizon ($\xi\to-\infty$). In fact, such a dependence on coordinates follows from the tensor structure. Consider a stress energy tensor of the form $\langle T^{\mu}_{\nu} \rangle = f(\xi)\text{Diag}(-3,1,1,1)$ in the Rindler coordinates. This should be covariant conserved quantity:
\begin{align}
  D_\mu  \langle T^{\mu}_{\nu}\rangle= \frac{\partial  \langle T^{\mu}_{\nu}\rangle}{\partial x^{\mu}}+ \langle T^{\alpha}_{\nu}\rangle\Gamma^\mu_{\alpha \mu}- \langle T^{\mu}_{\alpha}\rangle\Gamma^\alpha_{\nu\mu}=0.
\end{align}
The latter equation is a constraint on $f(\xi)$, that requires it to be proportional to the exponential factor $f(\xi)\propto \exp(-4\xi)$. This dependence on the coordinates corresponds to the strong effect of quantum fluctuations near the horizon. Such large fluctuations should deform the background metri, see 
\cite{Ho:2018jkm,Ho:2018fwq} and also \cite{Baake:2023gxx}.

\subsection{Loop correction in higher dimensions }
\label{Highdim}
In this section, we consider the case of an arbitrary space-time dimension. The regularized bare propagator at coincident points is as follows:
\begin{gather}
    G^{reg}(X,X)=  \ 
\raisebox{-.4\height}{\includegraphics[scale = 0.9]{graph_debye_10.pdf}}  \ = e^{-(d-2) \xi }
\times\\ \times \nonumber
\frac{1  
   }{
 2^{d-2} \pi ^{\frac{d+1}{2}} \Gamma \left(\frac{d-1}{2}\right)}\int_0^\infty d\omega \sinh (\pi  w) \Big|\Gamma \left(\frac{d-2}{2}-i \omega\right)\Big|^2 \left(\frac{1}{e^{\beta  \omega}-1}-\frac{1}{e^{2 \pi  \omega}-1}\right).
\end{gather}
Here we treat the UV divergent terms similarly as in four dimensions. As one can see, the behavior of the propagator at the coincident points on the horizon essentially depends on the dimension of the space. It gets worse in higher dimensions. On the other hand, the square root of the determinant of the metric does not change. This leads to the fact that the following volume integral in the loop:
\begin{align}
\int_{-\infty}^{\infty}  \frac{d \xi}{e^{(d-4)\xi}} K_{i \omega}(k e^{\xi} )K_{i p }(k e^{\xi}),
\end{align}
contains a divergence at the horizon, i.e. as $\xi\to -\infty$. In the case $d=4$ , this integral reduced to the delta function. As the result in $d=4$ there are no any problems of the type that we encounter then $d \ne 4$. In the higher dimensional case, after treating of the UV divergent terms at the coincident points of the propagator, one should also regularize the volume integral. 

Let us consider the following integral, which converges only for $d<4$:
\begin{align}
\label{deltad}
\int_{-\infty}^{\infty}  \frac{d\xi}{e^{(d-4)\xi}} K_{i \omega}(k e^{\xi} )K_{i p }(k e^{\xi})
= k^{d-4}  \frac{2^{1-d} \Bigg|\Gamma \left(\frac{4-d}{2}+\frac{1}{2} i (p-w)\right)\Bigg|^2 \Bigg| \Gamma \left(\frac{4-d}{2}+\frac{1}{2} i (p+w)\right)\Bigg|^2}{\Gamma (4-d)}.
\end{align}
Here we can use the dimensional regularization and, then, analytically continue the function of $d$ to an arbitrary even dimensions.  For $\omega \ne p$ the RHS of \eqref{deltad} vanishes for integer $d \geq 4$ due to the gamma function at the denominator. At the same time in the limits $\omega\to p$ and $d = n - \epsilon $, $\epsilon \to 0$ one gets:
\begin{gather}
\Bigg[\int_{-\infty}^{\infty}  \frac{d \xi}{e^{(d-4)\xi}} K_{i \omega}(k e^{\xi} )K_{i p }(k e^{\xi})\Bigg]_{reg}
= k^{d-4}  \frac{\Gamma(d-3) }{2^{d-3}\Gamma \left(\frac{d-2}{2}\right)^2}\Bigg| \Gamma \left(\frac{4-d}{2}+ i\frac{\omega+p}{2} \right)\Bigg|^2  \lim_{\epsilon \to 0}\frac{ \epsilon}{ \epsilon^2+(\omega-p)^2}
\nonumber
=\\=
k^{d-4}  \frac{\pi\Gamma(d-3) }{2^{d-3}\Gamma \left(\frac{d-2}{2}\right)^2}\Bigg| \Gamma \left(\frac{4-d}{2}+ i\omega \right)\Bigg|^2  \delta(\omega-p).
\end{gather}
In this way we get the orthogonality relation for the MacDonald functions in the higher dimensional case. As in the case of four dimensions this helps us to evaluate the sum of the subclass of diagram shown on the Fig.\ref{fig1}. Straightforward calculations gives the following expression:
\begin{gather}
\label{qwe}
    G(X_1,X_2)
    =\\= 
    \nonumber
    \frac{1}{\beta}\sum_{\omega_n}\int \frac{d^{d-2} \vec{k}}{(2 \pi)^{d-2}} \int_0^{\infty} \frac{ d \omega}{\pi^2} \frac{2\omega\sinh\pi \omega}{\w_n^2+\omega^2+P^2(\omega,k)}e^{-i\w_n (\eta_2-\eta_1)}e^{i\vec{k}(\vec{z}_2-\vec{z}_1)}K_{i \omega}\big(ke^{\xi_1}\big)K_{i\omega}\big(ke^{\xi_2}\big),
\end{gather}
where:
\begin{gather}
P^2(\omega,k)
= \lambda k^{d-4}  \omega \sinh(\pi \omega)\Bigg| \Gamma \left(\frac{4-d}{2}+ i\omega \right)\Bigg|^2 
\times
\\ 
\times
\nonumber
\frac{2^{2-d} \pi ^{-\frac{d}{2}-2}}{(d-3) \Gamma
   \left(\frac{d-2}{2}\right)}
\int_0^\infty d p\sinh (\pi  p) \Big|\Gamma \left(\frac{d-2}{2}-i p\right)\Big|^2 \left(\frac{1}{e^{\beta  p}-1}-\frac{1}{e^{2 \pi  p}-1}\right).
\end{gather}

As one can see, in higher dimensions for $\lambda \phi^4 $ theory the dressed propagator has a more complicated structure than in four dimensions. In fact, the integrand in \eqref{qwe} contains an additional dependence on $k$ and $\omega$. For example for the case $d=6$:
\begin{align}
P^2(\omega,k)=\frac{\lambda}{11 520 \pi^3} \frac{k^2}{\omega^2+1}\left[ \left(\frac{2\pi}{\beta}\right)^4+ 10  \left(\frac{2\pi}{\beta}\right)^2 -11  \right].
\end{align}
This dependence appears due to the fact that we had to use dimensional regularization to regularize both the space and momentum integrals. And it appears due to the fact that beyond $d=4$ the coupling constant $\lambda$ is dimensionfull and the theory under consideration is not conformal anymore.
\section{Conclusions and dicsussion}
In this paper, we discuss conformal scalar  quantum field theory in the Minkowski and Rindler coordinates with finite temperature planckian distribution for exact modes. Quantum loop effects in the self interacting $\lambda\phi^4$ theory lead to the generation of a thermal (Debye) mass. We make the following observations:
\begin{itemize}
\item In the Minkowski coordinates the thermal mass enters the propagator in the same place as the physical mass. Thus, in the spectrum these two masses (physical and thermal) look similar. Namely thermal (Debye) mass creates a mass gap in the spectrum. 

\item Moreover, in the Minkowski coordinates, the trace of the stress–energy tensor depends on the physical mass. But, at least in the first and 3/2 orders in the expansion over the coupling constant, the tensor remains traceless in presence of the thermal mass. The non-zero contribution to the trace coming from the term $\lambda g_{\mu\nu} \varphi^4$ in \eqref{genteimin} cancels the non-zero contribution to the trace coming from the terms with derivatives. This is in agreement with the fact that in flat space-time the stress-energy tensor remains traceless even on the quantum level. 

\item  At the same time, in the Rindler coordinates, the thermal mass enters the propagator differently from the physical mass. Thermal loops lead to the manifestation of the mass gap in the spectrum, while the physical mass does not do that. Thus, the thermal mass and the physical mass have different physical revelations in the Rindler coordinates. For the both type of  coordinates the thermal mass is not an analytical function of $\lambda$.

\item It also turns out that in the Rindler coordinates the corrections to the stress energy tensor are similar to the corrections in the Minkowski coordinates. Namely, in Rindler coordinates we also obtain traceless stress-energy tensor, despite the fact that mass is generated in the loops. 
\item Near the horizon, the stress energy tensor in the Rindler coordinates grows infinitely, which indicates the need to take into account the back reaction on the background geometry.

\item  We also show that in the Rindler coordinates the theory is self-consistent only if the temperature is greater than or equal to the canonical (Unruh) temperature. Otherwise a tachyon appears in the spectrum. 

\item We also discuss loop corrections to the propagator in higher dimensional space-time in Rindler coordinates. It is shown that in the case of $d\ne 4$ an additional dimensional regularization should be used to make the loop contributions well-defined. But the dressed propagator is more complicated than in four dimensional case, since the integrand of the dressed propagator contains an additional dependence on the energy and transverse momentum. That is due to the fact that in $d\ne 4$ the self-coupling constant $\lambda$ is dimensionfull. 
\end{itemize}
An area of the possible future research contains several topics:

It is interesting to consider loop corrections to the massive scalar field theory in the Rindler coordinates: thermal corrections to the trace of the stress-energy tensor and thermal corrections to the physical mass in Rindler coordinates.

In  non-stationary situations, such as time-dependent backgrounds, free Hamiltonian depends on time. This can lead to particle creation processes, see for example \cite{Akhmedov:2015xwa, Trunin:2021lwg}. In such a case, the dynamics of the theory strongly depends on the choice of the initial Cauchy surface, the basis of modes, and the choice of the initial state (for more details see: \cite{ Polyakov:2009nq, Polyakov:2012uc, Krotov:2010ma, Akhmedov:2019cfd, Akhmedov:2019esv, Akhmedov:2021rhq, Sadekov:2021rpy}). The future Rindler wedge is a toy model of such a situation \cite{Akhmedov:2021agm}. So it will be interesting to consider the same problems as in this article in the future non-stationary Rindler wedge, to find out how the dynamic of the dressed theory differs from the bare one.

Throughout this article we use Matrusbara technique. But, the similar calculations as in the paper can be performed using the Keldysh technique. Or even in 3x3 Keldysh technique. However, in some cases it corresponds to different physical situations \cite{Arseev:2015,Melo:2021mbd}. 

\section*{Acknowledgments}
We would like to acknowledge valuable discussions with A.G.Semenov and D.V. Fursaev. Especially we would like to thank E.T.Akhmedov for valuable discussions, sharing his ideas and correcting the text. This work was supported by the grant from the Foundation for the Advancement of Theoretical Physics and Mathematics ``BASIS'', by the Euler grant from the Saint Petersburg Leonhard Euler International Mathematical Institute and supported by the Ministry of Science and Higher Education of the Russian Federation (agreement no. 075–15–2022–287).
\newpage
\begin{appendices} 
\setcounter{equation}{0}
\renewcommand\theequation{A.\arabic{equation}}

\section{Usefull integrals \label{appA} }
\numberwithin{equation}{section}
In the this Appendix we give a short table of integrals which are used throughout this paper. These integrals can be found in e.g. \cite{Gradshteyn:1702455} or calculated using properties and integral representations of elementary and special functions \cite{NIST:DLMF,Akhmedova:2019bau}. Throughout this appendix it is assumed that:
\begin{align*}
\text{Im} \ M=0, \qquad M\ge 0, \qquad \text{Im} \ n=0, \qquad n>0;
\end{align*}
$K_{i \omega}(k e^{\xi}), I_1(2 M | \rho| ), Y_0( k a)$ and $\pmb{L}_{-1}(2 M \rho)$ are Bessel functions and modified Struve functions, correspondingly. Thus, we use the following integrals:
\begin{align}
\label{fourier}
\int_{-\infty}^{\infty} d \omega e^{2 i \omega \rho} \frac{|\omega|}{\sqrt{\omega^2+M^2}} = \pi  \delta (\rho)+\pi  M I_1(2 M | \rho| )-\pi  M \pmb{L}_{-1}(2 M \rho), \quad \text{Im} \ \rho=0
\end{align}
\begin{align}
\label{kintK}
\int_{0}^{\infty} d k k^{d-3} K_{i \omega}(k e^{\xi})K_{i \omega}(k e^{\xi})
=
e^{(2-d) \xi }\frac{\sqrt{\pi }  \Gamma \left(\frac{d-2}{2}\right) }{4 \Gamma   \left(\frac{d-1}{2}\right)} \Bigg|\Gamma   \left(\frac{d-2}{2}-i w\right)\Bigg|^2, \quad \text{Re}(\text{d})>2
\end{align}  
\begin{align}
\label{Y0}
\int_0^{\infty} dk k^{d} Y_0( k a)= \frac{2^d}{a^{d+1}} \tan \left(\frac{\pi d}{2}\right) \frac{\Gamma\left( \frac{1+d}{2}\right)}{\Gamma\left( \frac{1-d}{2}\right)}, \quad -1<\text{Re}(\text{d})<\frac{1}{2} 
\end{align}
\begin{align}
\int_0^\infty d \rho \frac{e^{- \alpha \rho}}{\sinh ^{\text{d}}(\rho)} = \frac{2^{d-1} \Gamma (1-d) \Gamma
   \left(\frac{d+\alpha}{2}\right)}{\Gamma \left(\frac{1}{2}
   (-d+\alpha+2)\right)}, \quad \text{Re}(\text{d})<1 \ \ \text{and} \ \ \text{Re}(\text{d}+\alpha)>0
\end{align}
Calculating derivatives with respect to $\alpha$ of the last integral and setting it to zero, one can obtain that: 
\begin{align}
\label{1/sinh}
\int_0^\infty d \rho \frac{ 1}{\sinh ^{\text{d}}(\rho)}=\frac{\Gamma \left(\frac{1-d}{2}\right) \Gamma
   \left(\frac{d}{2}\right)}{2 \sqrt{\pi }}, \quad \text{Re}(\text{d})<1
\end{align}
\begin{align}
\label{g/sinh}
\int_0^\infty d \rho \frac{ \rho}{\sinh ^{\text{d}}(\rho)}=\frac{1}{4} \sqrt{\pi } \cot \left(\frac{\pi  d}{2}\right) \Gamma
   \left(\frac{1-d}{2}\right) \Gamma
   \left(\frac{d}{2}\right), \quad \text{Re}(\text{d})<2
\end{align}
\begin{align}
\label{g^2/sinh}
\int_0^\infty d \rho \frac{ \rho^2}{\sinh ^{\text{d}}(\rho)}=\frac{\Gamma \left(\frac{1-d}{2}\right) \Gamma
   \left(\frac{d}{2}\right) \left[\pi ^2 \cot ^2\left(\frac{\pi 
   d}{2}\right)-\psi ^{(1)}\left(\frac{1-d}{2}\right)+\psi
   ^{(1)}\left(\frac{d}{2}\right)\right]}{8 \sqrt{\pi }}, \quad \text{Re}(\text{d})<3
\end{align}
We also need a series expansion in M of the following integrals:
\begin{align}
\label{J2pi}
     \int_0^\infty\frac{d\w \w^2}{\sqrt{\w^2+M^2}}\frac{1}{e^{2\pi\w}-1}=\frac{1}{24}-\frac{M}{2 \pi }+\frac{1}{8} M^2 \left(-2 \log \left(\frac{M}{2}\right)-2 \gamma -1\right)+\frac{\pi  M^3}{9}+O(M^4)
\end{align}
\begin{align}
\label{J4pi}
   \int_0^\infty\frac{d\w \w^4}{\sqrt{\w^2+M^2}}\frac{1}{e^{2\pi\w}-1}=\frac{1}{240}-\frac{M^2}{48}+\frac{M^3}{3 \pi }+O(M^4)
\end{align}
\begin{align}
\label{J2beta}
    \int_0^\infty\frac{d\w \w^2}{\sqrt{\w^2+M^2}}\frac{1}{e^{\beta\sqrt{\w^2+M^2}}-1}=\frac{\pi ^2}{6 \beta^2}-\frac{\pi  M}{2 \beta}+\frac{1}{8} M^2 \left(2 \log \left(\frac{4 \pi }{\beta
   M}\right)-2 \gamma +1\right)+O(M^4)
\end{align}
\begin{align}
\label{J4beta}
    \int_0^\infty\frac{d\w \w^4}{\sqrt{\w^2+M^2}}\frac{1}{e^{\beta\sqrt{\w^2+M^2}}-1}=\frac{\pi ^4}{15 \beta^4}-\frac{\pi ^2 M^2}{4 \beta^2}+\frac{\pi  M^3}{2 \beta}+O(M^4)
\end{align}
To clarify the results for \eqref{J2pi},\eqref{J4pi},\eqref{J2beta},\eqref{J4beta}, consider the following integral:
\begin{align*}
    \int_0^\infty\frac{d\w \w^n}{\sqrt{\w^2+M^2}}\frac{1}{e^{2\pi\w}-1}.
\end{align*}
We are interested in its expansion over the powers of $M$, as $M\to 0$. However, this integral is not an analitic function of $M^2$. Indeed, if we would naively use the Taylor expansion of $\left(\w^2+M^2\right)^{\frac{1}{2}}$ some of the integrals in the expansion will diverge. But, the integral under consideration can be evaluated using the Mellin transformation techniques:
\begin{multline*}
    \int_0^\infty\frac{d\w \w^n}{\sqrt{\w^2+M^2}}\frac{1}{e^{2\pi\w}-1}=\int_0^\infty\frac{d\w \w^n}{\sqrt{\w^2+M^2}}\sum_{m=1}^{\infty}e^{-m 2\pi \w}=\\=\int_0^\infty\frac{d\w \w^n}{\sqrt{\w^2+M^2}}\frac{1}{2\pi i}\int_{c-i\infty}^{c+i\infty}dz \sum_{m=1}^{\infty} \Gamma(z)(m 2\pi \w)^{-z}.
\end{multline*}
Where $c$ is some real number. Thus, the integral over $\w$ converges only for the case, when $n<c<n+1$. We choose $c=n+1/2$. Then:
\begin{align*}
    \int_0^\infty\frac{d\w \w^n}{\sqrt{\w^2+M^2}}\frac{1}{2\pi i}\int_{c+i\infty}^{c+i\infty}dz \sum_{m=1}^{\infty} \Gamma(z)(m 2\pi \w)^{-z} = \\ =\frac{1}{4\sqrt{\pi}\pi i}\int_{n+1/2-i\infty}^{n+1/2+i\infty}dz \sum_{m=1}^{\infty} \Gamma(z)(m 2\pi )^{-z} M^{n-z}\Gamma\Big(\frac{1+n-z}{2}\Big)\Big(\frac{-n+z}{2}\Big)=\\=
    \frac{1}{4\sqrt{\pi}\pi i}\int_{n+1/2-i\infty}^{n+1/2+i\infty}dz \zeta(z) \Gamma(z)( 2\pi )^{-z} M^{n-z}\Gamma\Big(\frac{1+n-z}{2}\Big)\Big(\frac{-n+z}{2}\Big).
\end{align*}
The latter integral can be calculated using the Cauchy residue theorem. The key point here, that in the small $M$ limit one shall look for the poles with the largest real part. The leading contributions come from $z=n,n-1,n-2,n-3$. Note, that we close a contour in the left half plane. This way we obtain \eqref{J2pi} and \eqref{J4pi}. The same idea can be used to calculate \eqref{J2beta} and \eqref{J4beta}.
\setcounter{equation}{0}
\renewcommand\theequation{B.\arabic{equation}}

\section{Dimensional regularization \label{appB}}
In this appendix we calculate the bare and dressed propagators and their derivatives at the coincident points with the use of the dimensional regularization. This is necessary for the calculation of the expectation value of the stress energy tensor.
\subsection*{The bare propagator \label{AppBarePropReg}}
The tree level Feynman propagator is:
\begin{gather}
\nonumber
    G_{\lambda^0}(X_1,X_2)=  \int \frac{d^2 \vec{k}}{(2 \pi)^2} \int_0^{\infty} \frac{ d \omega}{\pi^2} \sinh\pi \omega e^{i\vec{k}(\vec{z}_2-\vec{z}_1)}K_{i \omega}\big(ke^{\xi_1}\big)K_{i\omega}\big(k e^{\xi_2}\big)
    \times\\ \times
    \nonumber
  \left[e^{ -\omega (\eta_2-\eta_1)}\left(1+ \frac{1}{e^{\beta \omega }-1}\right)+e^{ \omega (\eta_2-\eta_1)}\frac{1}{e^{\beta \omega }-1}\right].
\end{gather}
For the canonical inverse temperature $\beta =2 \pi$ it is the standard Poincare invariant propagator:
\begin{align*}
    G_{\lambda^0}(X_1,X_2)\Big|_{\beta=2\pi} =\frac{1}{(2\pi)^2}\frac{1}{\sigma(X_1,X_2)},
\end{align*}
where $\sigma(X_1,X_2)$ is the squared geodesic distance. After the dimensional regularization the value of the massless Poincare invariant propagator at the coincident points is zero:
\begin{align*}
    G^{reg}_{\lambda^0}(X,X)\Big|_{\beta=2\pi}=0,
\end{align*}
because in the dimensional regularization the Poincare invariant propagator vanishes in the coincidence limit. Here the index $reg$ means that the regularization procedure is applied to the propagator. The last equality assumes that we have the following relation:
\begin{gather}
    \nonumber
  \bigg[   \int \frac{d^2 \vec{k}}{(2 \pi)^2} \int_0^{\infty} \frac{ d \omega}{\pi^2} \sinh\pi \omega  K_{i \omega}\big(ke^{\xi}\big)K_{i\omega}\big(k e^{\xi}\big) \bigg]^{reg}
   + \\ +
   \nonumber
  \int \frac{d^2 \vec{k}}{(2 \pi)^2} \int_0^{\infty} \frac{ d \omega}{\pi^2} \sinh\pi \omega K_{i \omega}\big(ke^{\xi}\big)K_{i\omega}\big(k e^{\xi}\big)  \frac{2}{e^{2 \pi \omega }-1} =0,
\end{gather}
where the first term is the regularized vacuum part of the propagator at the coincident points and the second contribution comes from the planckian distribution with the canonical temperature. Since for any temperature the vacuum contribution is the same, then the regularized bare propagator for any temperature at the coincident points acquires the following form:
\begin{gather}
\label{regbare}
       G^{reg}_{\lambda^0}(X,X) = \frac{e^{-2\xi}}{12}\Big(\frac{1}{\beta^2}-\frac{1}{(2\pi)^2}\Big).
\end{gather}
Here we use \eqref{kintK} to calculate the integral over $\vec{k}$ and \eqref{J2beta} to calculate the integral over $\omega$. One can see now that the propagator at the coincident points \eqref{regK} is proportional to $e^{-2\xi}$. Because of this factor we obtain the following orthogonality relation for the Bessel functions:  
\begin{align}
\int_{-\infty}^{\infty} d\xi  \ K_{i \omega}\big(ke^{\xi}\big)K_{i p}\big(ke^{\xi}\big) = \frac{\pi^2}{2}\frac{1}{\omega \sinh \pi \omega} \delta(\omega-p),
\end{align}
in the vertexes in the diagram. This delta function allows us to explicitly calculate contributions of the loops. 

Thus, in four dimensions the one loop correction to the propagator is as follows:
\begin{gather}
\triangle_1  G(X_1,X_2)=\
\ \raisebox{-.4\height}{\includegraphics[scale = 0.9]{graph_debye_2.pdf} }\
= -\frac{\lambda}{2}\int d V_{X} G_{\lambda^0}(X_1,X) G^{reg}_{\lambda^0}(X,X)G_{\lambda^0}(X,X_2) 
=  \\=
\nonumber
-\frac{1}{\beta}\sum_{\omega_n}\int \frac{d^2 \vec{k}}{(2 \pi)^2} \int_0^{\infty} \frac{ d \omega}{\pi^2} \frac{2\omega\sinh\pi \omega}{(\w_n^2+\omega^2)^2} e^{-i\w_n (\eta_2-\eta_1)}e^{i\vec{k}(\vec{z}_2-\vec{z}_1)}K_{i \omega}\big(ke^{\xi_1}\big)K_{i\omega}\big(ke^{\xi_2}\big)\frac{\lambda}{24}\Bigg(\frac{1}{\beta^2}-\frac{1}{(2\pi)^2}\Bigg). 
\end{gather}
This calculation heavily uses the fact that the regularized propagator at the coincident points is zero when $\beta=2\pi$. In the next section the eq. \eqref{regbare} will be obtained more rigorously via a different method.
\subsection*{The dressed propagator and its derivatives \label{AppDressedPropReg}}
The dressed propagator at the coincident points is (here we consider $M$ as just a parameter):
\begin{gather}
    G(X,X)=
    \frac{1}{\beta}\sum_{\omega_n}\int \frac{d^2 \vec{k}}{(2 \pi)^2} \int_0^{\infty} \frac{ d \omega}{\pi^2} \frac{2\omega\sinh\pi \omega}{\w_n^2+\omega^2+M^2}K_{i \omega}\big(ke^{\xi}\big)K_{i\omega}\big(ke^{\xi}\big).
\end{gather}
Since the propagator contains the term  $M^2$, we cannot relate it to the Poincaré invariant propagator and cannot use the same method as above in the Appendix \ref{AppBarePropReg}. So, we have to use the dimensional regularization explicitly, i.e. we will calculate the loop integral for generic value of number of dimensions $d$ for which the convergence is assured. Thus, let us consider the following integral:  
\begin{gather}
    G(X,X)
    \label{GGG}
    =\\= 
    \nonumber
    \frac{\mu^{4-d}}{\beta}\sum_{\omega_n}\int \frac{d^{d-2} \vec{k}}{(2 \pi)^{d-2}} \int_0^{\infty} \frac{ d \omega}{\pi^2} \frac{2\omega\sinh\pi \omega}{\w_n^2+\omega^2+M^2}K_{i \omega}\big(ke^{\xi}\big)K_{i\omega}\big(ke^{\xi}\big).
\end{gather}
where we introduced the arbitrary mass scale $\mu$ , in order to keep the correct units for the parameters and the integral. 

The evaluation of the sum over the Matsubara frequencies, gives:  
\begin{gather}
    G(X,X)
    =\\= 
    \nonumber
 \mu^{4-d}  \int \frac{d^{d-2} \vec{k}}{(2 \pi)^{d-2}} \int_0^{\infty} \frac{ d \omega}{\pi^2} \frac{\omega\sinh\pi \omega}{\sqrt{\omega^2+M^2}}K_{i \omega}\big(ke^{\xi}\big)K_{i\omega}\big(k e^{\xi}\big)\left(1+ \frac{2}{e^{\beta \sqrt{\omega^2+M^2} }-1}\right).
\end{gather}
Here, at $d=4$ the first term in the brackets has the UV divergence, while the second contribution is finite due to the Planck distribution. Since the first term depend on $M$ and it is UV divergent we should properly regularize it. Now let us add to and subtract from \eqref{GGG} the following term: 
\begin{align}
\mu^{4-d} \int \frac{d^{d-2} \vec{k}}{(2 \pi)^{d-2}} \int_0^{\infty} \frac{ d \omega}{\pi^2} \frac{\omega\sinh\pi \omega}{\sqrt{\omega^2+M^2}}K_{i \omega}\big(ke^{\xi}\big)K_{i\omega}\big(k e^{\xi}\big)  \frac{2}{e^{2 \pi w}-1}.
\end{align}
This way we get that:
\begin{gather}
\label{B8}
    G(X,X)
    =\\= 
    \nonumber
   \mu^{4-d}\int \frac{d^{d-2} \vec{k}}{(2 \pi)^{d-2}} \int_0^{\infty} \frac{ d \omega}{\pi^2} \frac{\omega\cosh\pi \omega}{\sqrt{\omega^2+M^2}}K_{i \omega}\big(ke^{\xi}\big)K_{i\omega}\big(k e^{\xi}\big)
       +\\+ 
    \nonumber
   \mu^{4-d}\int \frac{d^{d-2} \vec{k}}{(2 \pi)^{d-2}} \int_0^{\infty} \frac{ d \omega}{\pi^2} \frac{\omega\sinh\pi \omega}{\sqrt{\omega^2+M^2}}K_{i \omega}\big(ke^{\xi}\big)K_{i\omega}\big(k e^{\xi}\big)\left( \frac{2}{e^{\beta \sqrt{\omega^2+M^2} }-1}- \frac{2}{e^{2 \pi \omega }-1}\right).
\end{gather}
The second contribution on the RHS is finite for $d=4$, and it can be easily evaluated by taking the integrals over $k$ and $\omega$ using \eqref{kintK}, \eqref{J2pi} and \eqref{J2beta}:
\begin{multline}
    \mu^{4-d}\int \frac{d^{d-2} \vec{k}}{(2 \pi)^{d-2}} \int_0^{\infty} \frac{ d \omega}{\pi^2} \frac{\omega\sinh\pi \omega}{\sqrt{\omega^2+M^2}}K_{i \omega}\big(ke^{\xi}\big)K_{i\omega}\big(k e^{\xi}\big)\left( \frac{2}{e^{\beta \sqrt{\omega^2+M^2} }-1}- \frac{2}{e^{2 \pi \omega }-1}\right)=\\=e^{-2 \xi } \left(\frac{1}{12 }\left(\frac{1}{\beta^2}-\frac{1}{4 \pi ^2}\right)+M\left(\frac{1}{4 \pi
   ^3}-\frac{1}{4 \beta \pi }\right)+M^2\frac{ 1-\log \left(\frac{\beta}{2 \pi
   }\right)}{8 \pi ^2}-\frac{M^3}{18 \pi }+O(M^4)\right)
\end{multline}
To deal with the first contribution, we use the following integral representation \cite{Gradshteyn:1702455}:
\begin{align}
\cosh\pi \omega K_{i \omega}\big(ke^{\xi}\big)K_{i\omega}\big(k e^{\xi}\big) =-\pi \int_0^\infty d\rho \ Y_0 \left(2 k e^{\xi} \sinh \rho  \right) \cos(2 \omega \rho).
\end{align}
Substituting this expression into the first contribution on the RHS of \eqref{B8}, and taking the integrals with respect to $k$ and $\omega$, using \eqref{Y0} and \eqref{fourier}, one get:
\begin{gather}
\mu^{4-d}\int \frac{d^{d-2} \vec{k}}{(2 \pi)^{d-2}} \int_0^{\infty} \frac{ d \omega}{\pi^2} \frac{\omega\cosh\pi \omega}{\sqrt{\omega^2+M^2}}K_{i \omega}\big(ke^{\xi}\big)K_{i\omega}\big(k e^{\xi}\big)
= \\ =
\nonumber
\mu^{4-d}e^{(2-d) \xi } 2^{1-d} \pi ^{\frac{2-d}{2}} \cos \left(\frac{ \pi d }{2}\right)
   \Gamma \left(\frac{d-2}{2}\right)   
\int_0^\infty d \rho  \sinh ^{2-d}(\rho) \Big[  \delta (\rho)+M  
   \Big(I_1(2 \rho M)-\pmb{L}_{-1}(2 \rho M)\Big)\Big],
\end{gather}
where the $I $ and $\pmb{L}$  are modified Bessel and Struve functions, correspondingly. The first term in the bracket on the RHS is zero after the dimensional regularization. To evaluate the remaining terms we expand the integrand to the desired order in $M$ and take the integral over $\rho$ using \eqref{1/sinh},\eqref{g/sinh}. In the last step, we must expand the answer around $d=4$:
\begin{gather}
\mu^{4-d}e^{(2-d) \xi } 2^{1-d} \pi ^{\frac{2-d}{2}} \cos \left(\frac{ \pi d }{2}\right)
   \Gamma \left(\frac{d-2}{2}\right)   
\int_0^\infty d \rho  \sinh ^{2-d}(\rho) \Big[  \delta (\rho)+M  
   \Big(I_1(2 \rho M)-\pmb{L}_{-1}(2 \rho M)\Big)\Big]
   \nonumber 
   =\\=
   e^{-2 \xi } \left[-\frac{M}{4 \pi ^3}+M^2
   \left(\frac{1}{8 (-4+d) \pi ^2}-\frac{2+3 \gamma +\log (\pi )+2\log (\mu )+2 \xi
   }{16 \pi ^2}\right)+\frac{M^3}{18 \pi }\right] +O(M^4).
\end{gather}
In the end, both contributions together give the following answer:
\begin{gather}
\label{der0}
 G^{reg}(X,X)=\\=
 \nonumber
 e^{-2 \xi } \left[\frac{1}{12 }\left(\frac{1}{\beta^2}-\frac{1}{4 \pi ^2}\right)-\frac{M}{4  \pi\beta }+\frac{M^2}{8\pi ^2}
   \left(\frac{1}{ -4+d }-\frac{1}{2}\log \left(\frac{e^{2 \xi+3 \gamma}\beta^2  \mu^2 }{4 \pi
   }\right)
   \right)+O(M^4)\right]
\end{gather}
Now let us calculate derivatives of the dressed propagator at the coincident points.
\subsection*{Derivatives with respect to transverse directions}
The derivatives with respect to the transverse directions are:
\begin{gather}
  \partial_{\vec{z}_1}\partial_{\vec{z}_2}  G(X_1,X_2)\Big|_{X_1=X_2}=
    \frac{1}{\beta}\sum_{\omega_n}\int \frac{d^2 \vec{k}}{(2 \pi)^2} k^2 \int_0^{\infty} \frac{ d \omega}{\pi^2} \frac{2\omega\sinh\pi \omega}{\w_n^2+\omega^2+M^2}K_{i \omega}\big(ke^{\xi}\big)K_{i\omega}\big(ke^{\xi}\big).
\end{gather}
Thus, in this case, as compared to the one described above, the integrand contains an additional multiplier $k^2$, so one can do similar calculations to get:
\begin{multline}
\label{derz}
e^{4\xi}\partial_{\vec{z}_1}\partial_{\vec{z}_2}  G(X_1,X_2)\Big|_{X_1=X_2}= \frac{-11 \beta ^4+40 \pi ^2 \beta ^2+16 \pi ^4}{720 \pi ^2 \beta ^4}+\frac{\left(\beta -2 \pi ^2\right) M}{12 \pi ^3 \beta }+\\+\frac{M^2}{144 \pi ^2} \left(-\frac{12 \pi ^2}{\beta ^2}-6 \log \left(\pi ^2 \gamma ^{3/2}
   \mu  e^{2 \xi }\right)+\frac{6}{d-4}\right)+\frac{\left(\pi ^2 (\beta +9)-3 \beta \right) M^3}{54 \pi ^3 \beta }.
\end{multline}
\subsection*{Derivatives with respect to the coordinate along the acceleration}
This case is a little bit more tricky. The derivatives with respect to $\xi$ are:
\begin{gather}
  \partial_{\xi_1}\partial_{\xi_2}  G(X_1,X_2)\Big|_{X_1=X_2}=
    \frac{1}{\beta}\sum_{\omega_n}\int \frac{d^2 \vec{k}}{(2 \pi)^2}  \int_0^{\infty} \frac{ d \omega}{\pi^2} \frac{2\omega\sinh\pi \omega}{\w_n^2+\omega^2+M^2}
    \partial_{\xi} K_{i \omega}\big(ke^{\xi}\big)\partial_{\xi}K_{i\omega}\big(ke^{\xi}\big).
\end{gather}
Using the following relations:
\begin{align}
\partial_{\xi} K_{i \omega}\big(ke^{\xi}\big)\partial_{\xi}K_{i\omega}\big(ke^{\xi}\big)=\frac{1}{2}\partial^2_{\xi}\Bigg[ K_{i \omega}\big(ke^{\xi}\big) K_{i\omega}\big(ke^{\xi}\big)\Bigg]-  K_{i \omega}\big(ke^{\xi}\big)  \partial^2_{\xi} K_{i\omega}\big(ke^{\xi}\big)
\end{align}
and
\begin{align}
\partial^2_{\xi} K_{i\omega}\big(ke^{\xi}\big)=\left(k^2 e^{2 \xi }-\omega^2\right) K_{i \omega}\left(e^{\xi } k\right),
\end{align}
one can obtain that:
\begin{gather}
  \partial_{\xi_1}\partial_{\xi_2}  G(X_1,X_2)\Big|_{X_1=X_2}= \frac{1}{2} \partial^2_{\xi} G(X,X)-e^{2\xi}  \partial_{\vec{z}_1}\partial_{\vec{z}_2}  G(X_1,X_2)\Big|_{X_1=X_2}
  -\\ -
  \nonumber
      \frac{1}{\beta}\sum_{\omega_n}\int \frac{d^2 \vec{k}}{(2 \pi)^2}  \int_0^{\infty} \frac{ d \omega}{\pi^2} \frac{2\omega^3\sinh\pi \omega}{\w_n^2+\omega^2+M^2}
    K_{i \omega}\big(ke^{\xi}\big)K_{i\omega}\big(ke^{\xi}\big).
\end{gather}
where the first two terms on the RHS were calculated above. The third term can be calculated, using the same methods as above. As the result, one gets:
\begin{multline}
\label{derxi}
 e^{2\xi}\partial_{\xi_1}\partial_{\xi_2}  G(X_1,X_2)\Big|_{X_1=X_2}=\frac{\pi ^2}{90 \beta ^4}+\frac{1}{9 \beta ^2}-\frac{41}{1440 \pi ^2}+\\+\frac{M^2}{4320}\left(\frac{720}{\pi ^2 (d-4)}-\frac{180}{\beta ^2}-\frac{720 \log \left(\pi ^2 \gamma ^{3/2} \mu  e^{2 \xi
   +\frac{1}{16}}\right)}{\pi ^2}\right)-\frac{41}{1440 \pi ^2}\frac{(8 \beta +9) M^3}{108 \pi  \beta }-\frac{M}{3 \pi  \beta }.
\end{multline}
\subsection*{Derivatives with respect to time-like coordinate}
The derivatives with respect to $\eta$ contain an extra divergence from the sum over the Matsubara frequencies. But, this divergence does not depend on the thermal mass. Therefore, to deal with it we can use the result for the case $M=0$. The case $M=0$ is equivalent to the free theory, and the stress-energy tensor of the free theory was evaluated many times, e.g. \cite{Candelas:1977zza}.

As for the finite contributions, they can be calculated in the same way as in the previous subsections. Omitting the details we write down only the final answer:
\begin{align}
\label{dereta}
 e^{2\xi}\partial_{\eta_1}\partial_{\eta_2}  G(X_1,X_2)\Big|_{X_1=X_2}=\frac{\beta ^4-(2 \pi)^4}{480 \pi ^2 \beta ^4}+\frac{\left(\beta ^2+4 \pi
   ^2\right) M^2}{96 \pi ^2 \beta ^2}.
\end{align}

\section{Orthogonality relation of the MacDonald's functions  \label{appC}}
\renewcommand\theequation{C.\arabic{equation}}
Consider the following integral \cite{Gradshteyn:1702455}:
\begin{align*}
\int_{-\infty}^{\infty}  \frac{d\xi}{e^{(d-4)\xi}} K_{i \omega}(k e^{\xi} )K_{i p }(k e^{\xi})
= k^{d-4}  \frac{2^{1-d} \Bigg|\Gamma \left(\frac{4-d}{2}+\frac{1}{2} i (p-w)\right)\Bigg|^2 \Bigg| \Gamma \left(\frac{4-d}{2}+\frac{1}{2} i (p+w)\right)\Bigg|^2}{\Gamma (4-d)},
\end{align*}
which is convergent for $d<4$. For $\omega \ne p$, the integral does vanish in the limit  $4-d=\epsilon \to +0$ due to the gamma function in the denominator. Hence in the limits $\omega\to p$ and $4-d=\epsilon \to +0$, one obtains the orthogonality relation for the MacDonald's functions:
\begin{align}
 \int_{-\infty}^{\infty} d\xi  K_{i \omega}(k e^{\xi} )K_{i p }(k e^{\xi})
= \frac{1}{2}\Big|\Gamma(i\omega)\Big|^2\lim_{\epsilon \to 0}\frac{ \epsilon}{ \epsilon^2+(\omega-p)^2} = \frac{\pi^2}{2}\frac{1}{\omega \sinh \pi \omega} \delta(\omega-p).
\end{align}
This integral agrees with the answer obtained in \cite{PASSIAN2009380}.

\end{appendices}
\newpage
\bibliographystyle{unsrturl}
\bibliography{bibliography.bib}
\end{document}